# The Non-Ideal Organic Electrochemical Transistors Impedance

**Sébastien Pecqueur[1,§,*], Ivor Lončarić[2,§], Vinko Zlatić[2], Dominique Vuillaume[1], Željko Crljen[2,*]**

[1] Institute of Electronics, Microelectronics and Nanotechnologies (IEMN), CNRS, Univ. Lille, Villeneuve d'Ascq, France
[2] Ruđer Bošković Institute, P.O. Box 180, 10002 Zagreb, Croatia
[§] these authors have contributed equally
* E-mail: sebastien.pecqueur@*iemn.univ-lille1*.fr; zeljko.crljen@*irb*.hr

**Organic electrochemical transistors offer powerful functionalities for biosensors and neuro-inspired electronics, with still much to understand on the time-dependent behavior of this electrochemical device. Here, we report on distributed-element modeling of the impedance of such micro-fabricated device, systematically performed under a large concentration variation for $KCl_{(aq)}$ and $CaCl_{2(aq)}$. We propose a new model which takes into account three main deviations to ideality, that were systematically observed, caused by both the materials and the device complexity, over large frequency range (1 Hz to 1 MHz). More than introducing more freedom degree, the introduction of these non-redundant parameters and the study of their behaviors as function of the electrolyte concentration and applied voltage give a more detailed picture of the OECT working principles. This optimized model can be further useful for improving OECT performances in many applications (e.g. biosensors, neuro-inspired devices,…) and circuit simulations.**

## 1. Introduction

Organic electrochemical transistors (OECTs) are a class of iono-electronic devices that are demonstrating growing promises for both *sensing* and also for *emulating* neural activity.[1-5] The specific operating principle of this conducting-polymer based device makes it particularly sensitive to ions from an interfacing electrolyte medium,[6] and transduces this sensitivity to a transient electrical response, which can mimic neurons.[7,8] This growing interest prompts a better understanding of the transient behavior of these devices over the largest frequency range. Impedance spectroscopy studies on poly(3,4-ethylenedioxythiophene):poly(styrene sulfonate) (PEDOT:PSS) based OECTs have revealed two distinctive operation regimes for the device:[9] a low frequency regime at which the electrical properties are dominated by the conductance of the conducting polymer, and a higher frequency regime (typically for frequency “f” above 1 to 10 kHz) at which the device response is rather limited by the conductance of the gating electrolyte. It was shown that both frequency regimes involve two different ion-dependent physical processes and that each process can be dominant at two distinctive frequencies. More importantly, the nature of the cation has a specific effect on the impedance spectra with distinctive contributions on each of the two different frequency regimes. Consequently, the OECT can be used as a bi-parametric ion sensor, with both low and high frequency responses providing the necessary information to detect cations in a selective way.[9] Simple models have been used to simulate the frequency and transient response of OECT, based on a "1R1C" model (one resistor, one capacitor),[10] or a “2R1C" model (composed of two resistors and a capacitor),[9,11] considering a geometric capacitor characteristic of the electrolyte gate coupling, one resistor characteristic of the ion-dependent conduction through the polymer, and one resistor characteristic of the ion-dependent conductivity of the electrolyte. However these elementary models are not able to catch non-idealities that can be systematically identified on the impedance spectra, particularly at frequencies “f” above 100 kHz for low ionic concentrations. Indeed, experimental observations of concentration-dependent negative phase change in the impedance behavior suggests an inductive contribution to the device impedance under this regime, which is quantitatively ruled by the ionic low concentration.[9] Here, motivated by the understanding of such non-idealities, we report on the modeling of OECTs by an optimized equivalent circuit model taking into account the whole frequency dependency of the device impedance from 1 Hz to 1 MHz and harmonized for the largest ionic concentration range ($10^{-4}$ – 1 M) and various chemical nature of the ions. We also discuss the physical interpretation of the newly introduced circuit elements and on their contribution in the ion sensing of the device impedance. We quantitatively correlate the different electrolytes and voltage conditions with the individual circuit elements of our OECT model, which takes into account the practical non-idealities to the 2R1C model. These correlations evidence the local effect of cations at multiple levels in the OECT architecture responsible for the specific effects on the impedance at different frequency regimes, unravelling its multi-parametric ion sensitivity for high-dimensional ion-sensing.

## 2. Experimental results

Elementary circuit modeling is often a powerful mean to picture the electronic processes involved in a sophisticated device architecture. Therefore, the most representative fitting to the experimental device impedance leads to an optimal interpretation of the electronic-charge related mechanisms occurring upon its transient operation. On this matter, OECTs are particularly challenging platforms for many aspects. First, the active material is a conductive polymers: versatile with both molecular aspects (chemically specific) and material ones (morphology, crystallinity) ruling their electronic properties. Second, the transistor architecture is a three-electrode configuration where the numerous interfaces increase the model complexity. Third, the specific electrolyte-gating in OECTs occurring in the bulk of the organic material and involving ions (as charge carriers and matter) are driven by multiple processes, which cannot accurately be modelled by discrete ideal Ohmic resistors and electrostatic capacitors. Up to now, RC models supported a qualitative representation of the device functioning, up to the exploitation of the OECT as a bi-parametric ion sensor. Nevertheless, many non-idealities in the impedance data to the proposed model, covering broad ranges of frequencies, attracted our attentions. To identify them precisely, we studied under different conditions the deviation from ideality of the impedance data for the OECTs in the Bode and Nyquist plots, considering as a starting point the proposed 2R1C model (or "R|(R+C)") as a parallel circuit of one resistor with a serial resistor and capacitor (see Figures S1-S3 for all Nyquist representations). Bode and phase plots of the OECT exposed in the different electrolytes, as well as fabrication details, have been published elsewhere.[9] The OECT has a concentric source-drain geometry with an inner Pt-drain electrode radius of 50 µm and channel length of 20 µm with an annular outer Pt-source electrode wide of 10 µm. The choice for these dimensions has been based on the will to realize micrometric OECT devices while maximizing both low and high frequency impedance modulations.[9] Both electrodes are covered with a 16±5 nm thick PEDOT:PSS layer, with an Ag-wire used as a gate. To perform the bipolar impedimetric characterization, we characterized the OECT with the grounded gate, the signal, polarized at $V_{DC}+V_a*\sin(2\pi f(t)*t)$, at the drain electrode ($V_a$ = 50mV and $V_{DC}$ the direct component of the voltage) and the source pull-down to the ground with a shielded resistive load. Previous studies have shown that such a circuit configuration for the impedimetric study of an OECT dynamics allows the dual stimulation required for the transistor operation.[9] Ion charging promoted by the gate polarization and the hole drift promoted by the drain voltage can be simultaneously monitored at different frequencies in a single setup by the mean of a pull-down resistor. Using such a configuration involving a passive element instead of an external voltage generator offers the flexibility to modify voltage biases between source drain and gate without introducing further artefacts from the impedance of external active circuits.

2.a - High Frequency Non-Idealities

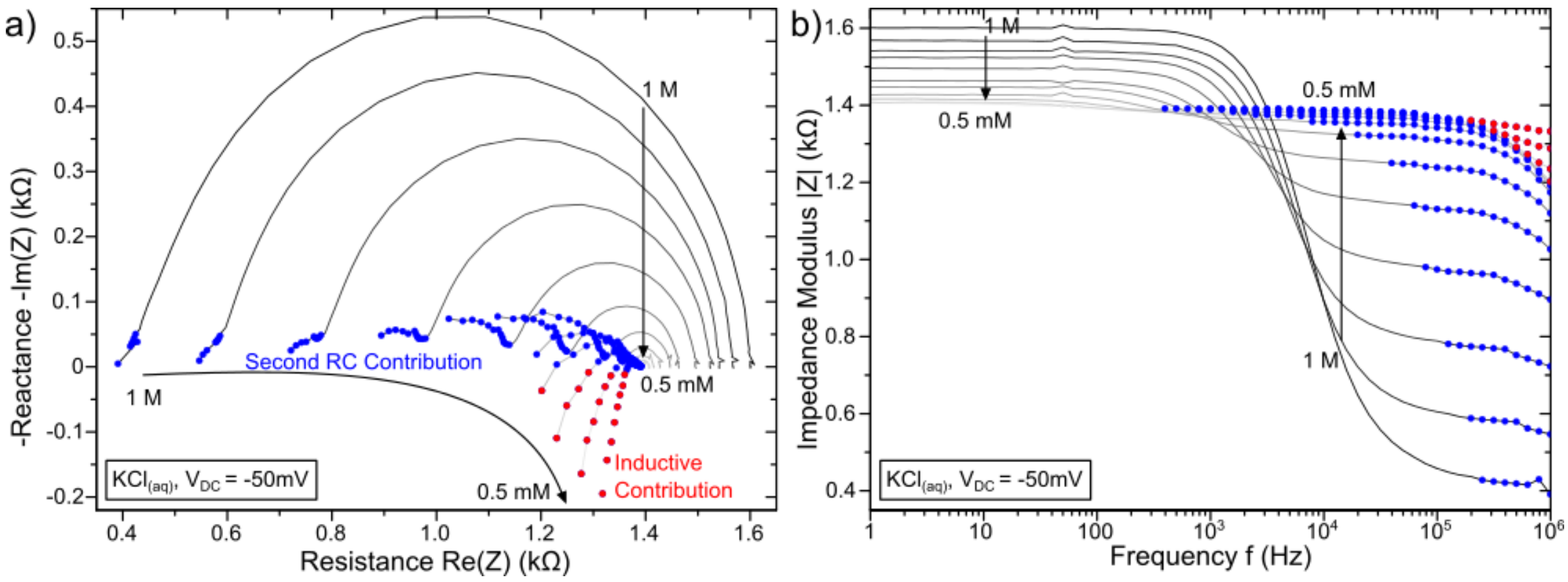

Figure 1 – Nyquist plot (a) and Bode impedance modulus representation (b) for the impedance spectroscopy of the concentration variation of $KCl_{(aq)}$ electrolytes (the arrow points out the impedance variation with the exponential-2 dilution of the electrolytes. Concentrations: 1 M, 0.5 M, 0.25 M, 125 mM, 62.5 mM, 31.3 mM, 15.6 mM, 7.81 mM, 3.91 mM, 1.95 mM, 0.98 mM and 0.49 mM: Polarization: $V_{DC}$ = -50 mV). The highlighted domains represent the main two deviations from the original 2R1C model at high frequency: The inductive contribution (red) displaying a positive reactance in Nyquist, and a second RC discharge (blue) displaying a second semicircle in Nyquist. Color highlights from both graphs are representative of the same experiment data, in order to appreciate their concentration-dependent contributions in the frequency domain.

Upon small voltage biases ($V_{DC}$ = -50 mV), the impedance data displays two specific concentration-dependent behaviors at high frequency (Figure 1). On the Nyquist plots, a first deviation “D1” to the 2R1C-model ideality is visible as a second semicircle characteristic from a second RC transition, which appears at any concentration (though more pronounced above 4 mM) and for any salts (Figures S1). The Bode representation shows that this feature affects a non-negligible part of the impedance spectrum for the high frequencies (typically above 1 to 100 kHz depending on the concentration). For the most diluted electrolytes (typically below 4 mM – Figure S1), the Nyquist plot displays a second deviation “D2” to the 2R1C-model ideality, characterized by a positive imaginary contribution, suggesting the presence of an inductor in the equivalent circuit model. This feature can also be identified on Bode phase plots (which have been published elsewhere)[9] by an increase of the impedance phase at low concentrations and high frequencies. The corresponding Bode representation of the impedance modulus shows that this impact is confined the high frequency range, typically above 100 kHz.

2.b - Low Frequency Non-Idealities

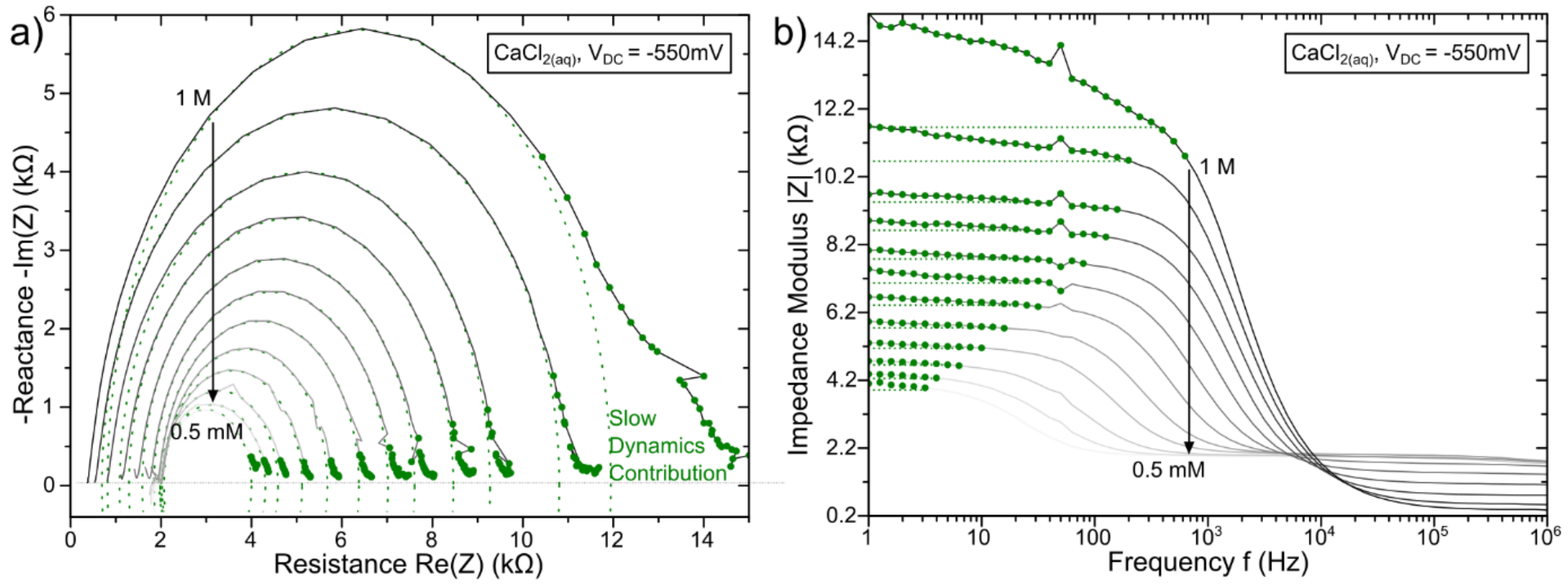

Figure 2 – Nyquist plot (a) and Bode impedance modulus representation (b) for the impedance spectroscopy of the concentration variation of $CaCl_{2(aq)}$ electrolytes (the arrow points out the impedance variation with the exponential-2 dilution of the electrolytes. Concentrations: 1 M, 0.5 M, 0.25 M, 125 mM, 62.5 mM, 31.3 mM, 15.6 mM, 7.81 mM, 3.91 mM, 1.95 mM, 0.98 mM and 0.49 mM: Polarization: $V_{DC}$ = -550 mV). The highlighted domain represent the main deviation from the original 2R1C model at low frequency: The dynamics contribution associated to a Constant Phase Element (green) displaying a deviation from an ideal Nyquist RC circle. Color highlights from both graphs are representative of the same experiment data, in order to appreciate their concentration-dependent contributions in the frequency domain.

For larger voltage biases ($V_{DC}$ = -550 mV), the PEDOT:PSS channel resistance contribution at low frequencies is enhanced compared to the expected plateau in a 2R1C model (Figure 2), which promotes non idealities related to the polymer dedoping (Figure 2). On the Nyquist plots, the first semicircle shows a third deviation “D3” on the low frequency domain (high resistance values) characteristic of a slow dynamics at frequencies up to 1 kHz for 1 M $CaCl_{2(aq)}$.

According to the three deviations to the 2R1C-model ideality that have been identified on the whole frequency spectrum of the characterized electrolytes, the necessity to increase the equivalent model complexity by the introduction of three elemental circuits has been highlighted.

### 3. Model

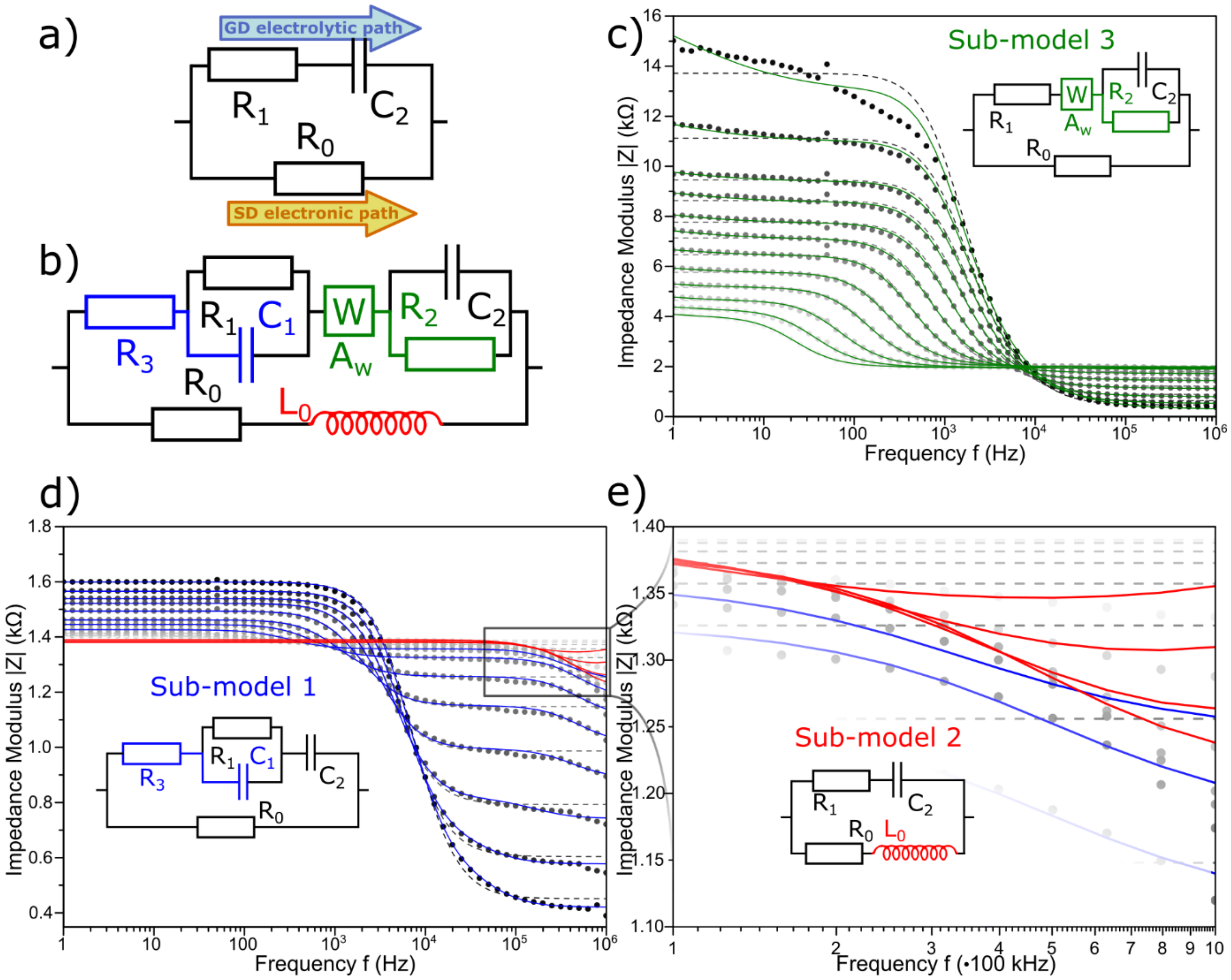

Figure 3 – a) Original "2R1C" model previously proposed,[9,11] to justify the concentration dependency of the OECT setup. b) Revised equivalent circuit with color highlights on the different electrical elements proposed to fit the three different non-idealities D1 (blue), D2 (red) and D3 (green). External resistive contributions as well as positions of the source, drain and gate contacts are further depicted in the Figure S4 as supplementary information. Impedance modulus data (circles) of the previously (Figure 2) displayed $CaCl_{2(aq)}$ at $V_{DC}$ = -550 mV (c) and $KCl_{(aq)}$ electrolytes (Figure 1) at $V_{DC}$ = -50 mV (d,e) and fitting over the whole range with the original 2R1C model (dashes) and with the optimized reduced models (colored lines, equivalent circuits as insets).

From the previous analysis of the impedance spectrum over the whole range of frequency and the different electrolytic environments, one can conclude on the necessity to add the following elements to the original 2R1C model (displayed in figure 3a): a resistor and a capacitor for the simulation of D1, an inductor for the simulation of D2 and a constant phase element for D3. In order to determine the most relevant position of these new elements within the optimized equivalent circuit, we identified their contribution domination at the specific frequency domains, considering the 2R1C equivalent circuit and the physical significance of the two resistors and one capacitor original elements ($R_0$, $R_1$ and $C_2$ in figure 3b, where $R_0$ is the equivalent resistive contribution of the source-drain (SD) electronic conduction, $R_1$ is the equivalent resistive contribution of the gate-drain (GD) electrolytic conduction, and $C_2$ is equivalent capacitive contribution of the electrolyte/electronic coupling). At this stage, the introduction of the new terms is strictly based on the needs of flexibility to optimize the fits for all impedance data, reduced to the stringent number of elements to avoid redundant flexibility, and arranged in the unique way for fitting at best and evenly any frequency-dependent deviation for any electrolyte. This approach is comparable to the circuit deconstruction of an unknown system via signal processing by reverse-engineering. Therefore at this stage, no assumption based on the hypothetical physical significance of the newly introduced parameters are made to avoid biasing their interpretation in our analysis.

The first deviation D1 introduces a second semicircle on the Nyquist plot, which is translated on the Bode plot by a decrease of the impedance modulus at high frequency. This behavior characterizes a second capacitive coupling $C_1$ enabling a higher conduction through the system at a higher frequency. Noticing that the characteristic time of the second transition seems to increase with the decrease of the concentration (the blue transition shifting to lower frequencies in Figure1a), $C_1$ should be involved in a capacitive coupling shorting a resistor for which its resistance value increases with the decrease of the concentration. We can conclude that $C_1$ shorts the electrolytic GD path in parallel with $R_1$ (for which its value increases when decreasing the electrolyte concentration) rather than the electronic SD path in parallel with $R_0$ (for which its value decreases when decreasing the electrolyte concentration). Since the resistive component Re(Z) of the impedance at high frequencies after the second RC contribution does not converge to zero (Figure 1a), one can conclude on the existence of a residual electrolytic resistance $R_3$ which is not shortened by $C_1$ (see sub-model 1, figure 3d).

The second deviation D2 displays an inductive contribution by a positive reactance at high frequencies. Considering the fact that this positive reactance expresses only for devices with low-concentration electrolytes, this behavior suggests an inductor to be involved on the electronic SD path in series with $R_0$ (for which the resistance decreases with the electrolyte dilution, such as to become smaller than the inductor's impedance $L_0\omega$) rather than on the electrolytic GD path (for which the resistance increases with the electrolyte dilution, see sub-model 2, figure 3e).

The third deviation D3 expresses an imperfect behavior (depressed semicircle) at low frequencies involving the non-ideal impedance of a constant phase element (CPE) in parallel with a resistor. The increase of its expression with the increase of the concentration (Figure 2b) suggests this element to be on the electrolytic GD path for which the impedance decreases with the increase of concentration (and therefore $R_0$ being the parallel resistive contribution for the depressed semicircle). Since the expression of this constant phase element occurs before the first RC impedance drop, it suggests the capacitance $C_2$ to be shortened by a resistor $R_2$. To rigidify the model, we assumed this constant phase element to be a Warburg element for which the phase is fixed to 0.5 (the relevance of this element will be discussed later, see sub-model 3, figure 3c).

To verify the validity of the revised equivalent circuit, the systematic fitting of the data with reduced optimized equivalent circuits have been made to appreciate the contribution of the individual added features on the observed deviations (see figure 3c-e). Also, we studied the influences of these new elements have on the $R_0$, $R_1$ and $C_2$ contribution of the original 2R1C circuit in order to quantify their additive contribution to the original 2R1C model.

Figure 3d shows Bode plots for $KCl_{(aq)}$ ($V_{DC}$ = -50 mV, experimental data from figure 1a),and the fits with the 2R1C model for concentration from 0.5 mM to 1 M, as well as fits with the sub-model 1 (only non-ideality D1) for concentrations higher than 8 mM. The $R_0$ values are the same for both models (deviation below $10^{-4}$): the added features have no influences for low frequencies since both models converge to $R_0$. The sub-model 1 does not significantly modify the $C_2$ value (deviation below 3% of the fitted value for the 2R1C model), which is fixed by the frequency range between the two plateaus at low and high frequencies. The capacitance $C_1$ takes into account the beginning of the impedance decrease above 100 kHz, with $C_1$ being systematically 3-fold lower than $C_2$. In the sub-model 1, the $R_1$ value is increased by 50 to 70% of the original value obtained with the 2R1C model: one can observe on Figure 3d the deviation at high frequencies of the blue curves compared to the plateau of the 2R1C fitting for the corresponding data.

In Figure 3e and for $KCl_{(aq)}$ below 8 mM stimulated at $V_{DC}$ = -50 mV and although it particularly fits the phase increase at high frequencies, Sub-model 2 has a big impact on all three values since it does not take into account the main phase transition in the kHz range: $R_0$ is averaged between the fitted $R_0$ and $R_1$ value given by the 2R1C fitting while $C_2$ becomes characteristic from the MHz transition (with $C_1$ = 35±2 pF for $L_0$ = 29±5 µH). This shows that two independent capacitance are involved at high frequencies, with the necessity to add an inductive element for the MHz response at low concentrations.

On Figure 3c and for $CaCl_{2(aq)}$ electrolyte at $V_{DC}$ = -550 mV (for which the low frequency impedance is particularly deviating from the 2R1C model), the sub-model 3 modifies all three values since the additional features are added for fitting high impedance modulus deviations and affects the whole frequency range as they are not in series with any frequency-dependent element: nor capacitor or inductor. $R_0$ fitted values are increased by 15 to 65% while $R_1$ fitted values are decreased by 10 to 35%, compared to the values of the 2R1C fittings (as expected according to the relative position of $R_2$ in parallel with $R_0$ and in series with $R_1$). Because of the phase dependency of the Warburg element, $C_2$ values substantially lowered from 13% up to 117% for the 1 M concentration, compared to the value of the corresponding 2R1C fittings.

### 4. Results and Discussion

Before attempting to correlate the eight elements of the optimized model to the physical properties of the OECT, we first consider the external circuit element related to our setup (see Figure S4 in supplementary materials). As described previously,[9] the addition of a load resistor of 1 kΩ at the source was necessary to promote a sufficient capacitive gate coupling at high frequency (the PEDOT:PSS being a highly doped polymer with conductivities up to the kS·cm$^{-1}$ range,[12] one needs to balance the resistance of the electronic path with respect to the impedance of the electrolytic path). Additionally, contact lines of 46 Ω covered with Parylene C were necessary to contact the source and drain electrodes out of the electrolyte drop, in order to avoid capacitive coupling between the gate and the needle probes used to contact the source and the drain. Taking into account these constant resistive elements as part of the whole electrical circuit is perfectly compatible with our optimized model: $R_0$ is the resistive component of the electronic path's impedance, which embeds the resistance of both source and drain contact lines (2r) and the load ($R_{load}$), while $R_3$ is the only resistive contribution on the electrolytic path which is not in parallel to any capacitor and therefore shall embed the resistance of the drain contact line (r – see Figure S4 in supplementary materials for adapted equivalent circuits).

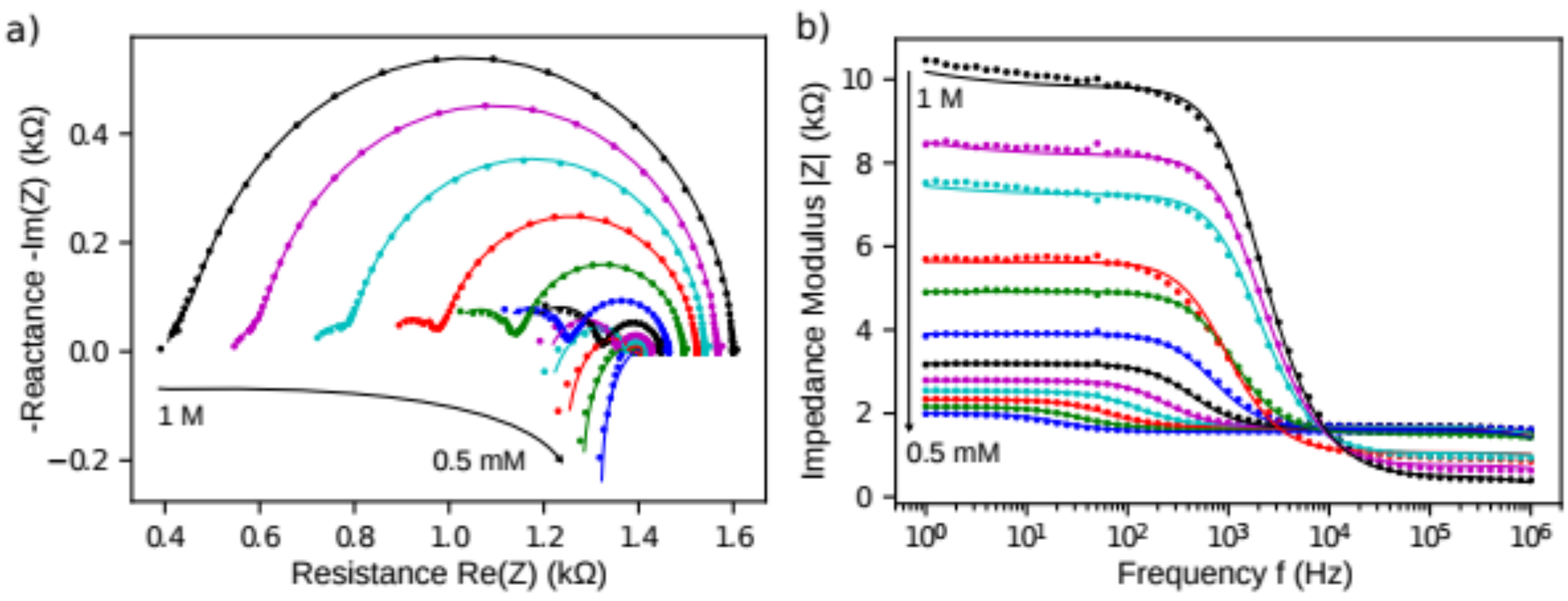

Figure 4 - Comparison of measurements and full model fits: a) Nyquist plot for $KCl_{(aq)}$ solutions at $V_{DC}$ = -50 mV, b) Bode impedance modulus for $KCl_{(aq)}$ solutions at $V_{DC}$= -550 mV.

In Figure 4, we show the measured spectra of $KCl_{(aq)}$ solutions for a set of concentrations compared to the fit by the full equivalent model introduced in previous section to account for the three non-idealities. For fitting the data, we have used Trust Region Reflective algorithm as implemented in SciPy,[13,14] constraining parameters in sensible ranges that are determined from simpler sub-models. The reason is that full model has eight circuit parameters that are not trivial to determine for all concentrations/voltages since the second RC contribution, the inductive contribution, and the slow dynamics contribution are not visible in all experimental conditions. Our starting point is always 2R1C model for which it is easy to fit parameters for the data in the regions without non-idealities. Constraining these parameters, we move to fit parameters corresponding to different non-idealities present in the data. In some cases non-idealities are not visible in the data, e.g. there is no inductive contribution at concentrations of 6 mM and higher, and therefore, reliable fitting for the corresponding parameter cannot be performed. In these cases we have constrained value of the parameters by hand (e.g. we have chosen $L_0$ = 0 at concentrations of 6 mM and higher).

### 4.1 – The First Deviation – The Second RC Contribution

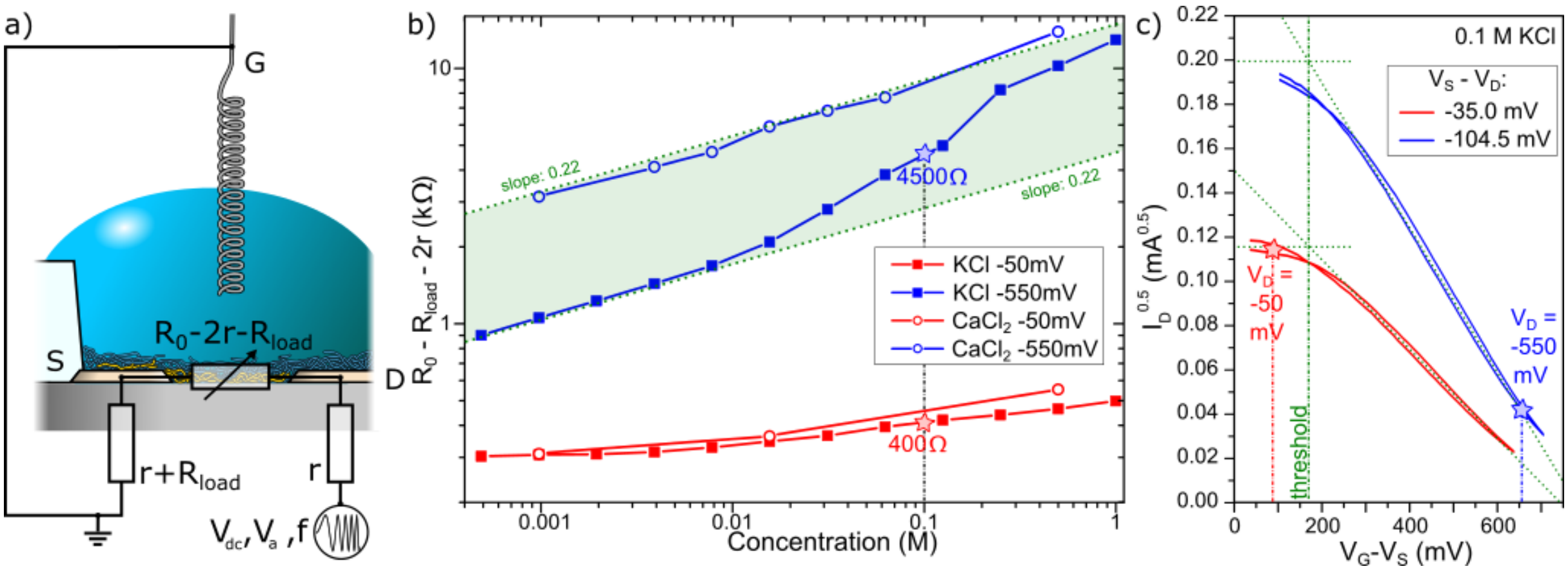

Figure 5 – (a) Experimental setup (omitting all frequency dependent terms) setting the steady component of the source potential. (b) Dependency of the fitted $R_0$-2r-$R_{load}$ values (parameters were fitted to the data with the model in Figure 3b). (c) Transfer characteristics (trace and retrace) performed in a 0.1 M $KCl_{(aq)}$ electrolyte and at two different source-drain voltages (gate wire grounded). The blue stars and red stars in Figures 5b and 5c correspond to two equivalent voltage and electrolyte conditions, showing that the transistor operates at two different regimes upon impedance spectroscopy.

In our setup, as the drain and the gate electrodes are directly polarized to, respectively, the signal and the ground, the DC component of the source electrode potential is set at an intermediate potential by a voltage divider between the drain potential, dropping through the channel resistance and the gate potential dropping through the load (Figure 5a). Therefore as it directly controls source potential (and therefore indirectly setting the source-drain voltage), it is particularly important to control the resistance of the PEDOT:PSS material and its cation sensitivity which controls the ion accumulation profile over the OECT. Its resistance ($R_0$ - 2r - $R_{load}$) increases with both the increase of the ion concentration and $V_{DC}$ (which controls the source-drain and gate-drain biases) from as few as 300 Ω up to the 10 kΩ range for 1 M-scale concentrated electrolytes (two orders of magnitude conductance modulation). Moreover, we observed the values not to be much sensitive to the nature of the electrolyte under low $V_{DC}$ bias, while $R_0$ appears to be more sensitive to calcium ions than to potassium ones at $V_{DC}$ = 550 mV (with a specific calcium over potassium resistance ratio of 2.9±0.1 for concentrations lower than $1.6 \cdot 10^{-2}$M): at low concentrations, three $K^+_{(aq)}$ affects the resistance as one $Ca^{2+}_{(aq)}$, showing that valency is not the single parameter. We also noticed that the trend for the resistance to increase with the concentration a slope much lower than one in the double logarithmic plot, which reveals that if neglecting hole injection and hole mobility dependency, the cation density increase in the electrolyte is not proportional to the decrease in hole density in the PEDOT:PSS. The strong coupling between the electronic properties at the bulk of the polymer and the electrolytic properties of the environment separating it from the gate is at the origin of the sensing mechanism for PEDOT:PSS. To support the p-doped conducting polymer reduction (removing holes) electrically induced by a negative voltage bias with respect to the gate potential, the electrolyte must provide mobile cations to satisfy locally the electro-neutrality in the material. The dedoping diminishes both hole transport and injection, due to PEDOT:PSS conductance decrease, as well as the contact resistance potentially associated to Schottky barriers,[15] which both additively contributes to the decrease of drain current of holes. Both effects are embedded in an equivalent resistor which increases in resistance with the increase of the electrolyte's ion concentration, which has been confirmed experimentally (Figure 5b). From the dimensions of our device and assuming transport limitation with no contact resistance (current assumed to be injected from the side-walls of the electrodes), the lowest resistance leads to the highest conductivity $\sigma_0$ for PEDOT:PSS of 124 S·cm$^{-1}$ which is an order of magnitude lower than the state-of-the art for a comparable compositions:[16] the possibility of partial contact limitation is therefore not excluded. To evaluate the impact of $V_{DC}$ on the operation of the transistor, the transfer characteristics of the device have been measured, (Figure 5c) at voltage conditions corresponding to the PEDOT:PSS resistance ($R_0$ - 2r - $R_{load}$) given by impedance spectroscopy (from Figure 5b). For the respective resistance values of 400 Ω ($V_{DC}$ =-50 mV) and 4500 Ω ($V_{DC}$ =-550 mV) in a 0.1 M $KCl_{(aq)}$ electrolyte, the corresponding DC component of the voltage at the source equals -35.0 mV (70% of the signal at $V_{DC}$ =-50 mV) and -104.5 mV (19% of the signal at $V_{DC}$ =-550 mV) according to the voltage divider displayed in Figure 5a. The transfer characteristics confirm that these two voltage conditions induce different driving modes in the transistor. When the drain is polarized at -50 mV/$V_G$ with a -35.0 mV drift voltage between the source and the drain (red stars in Figures 5b,c), the transistor is operated below the threshold voltage. While when the drain is polarized at -550 mV/$V_G$ with a -104.5 mV drift voltage between the source and the drain (blue stars in Figures 5b,c), the transistor is operated in the quadratic regime of the transfer characteristic. Considering that both regimes induce different dedoping profiles in the PEDOT:PSS semiconductor above the electrodes but also across the channel,[10] this reveals the pertinence of probing the OECT impedance at these two voltages, in order to extract non-redundant ion-dependent signals out of the PEDOT:PSS resistance ($R_0$ - 2r - $R_{load}$).

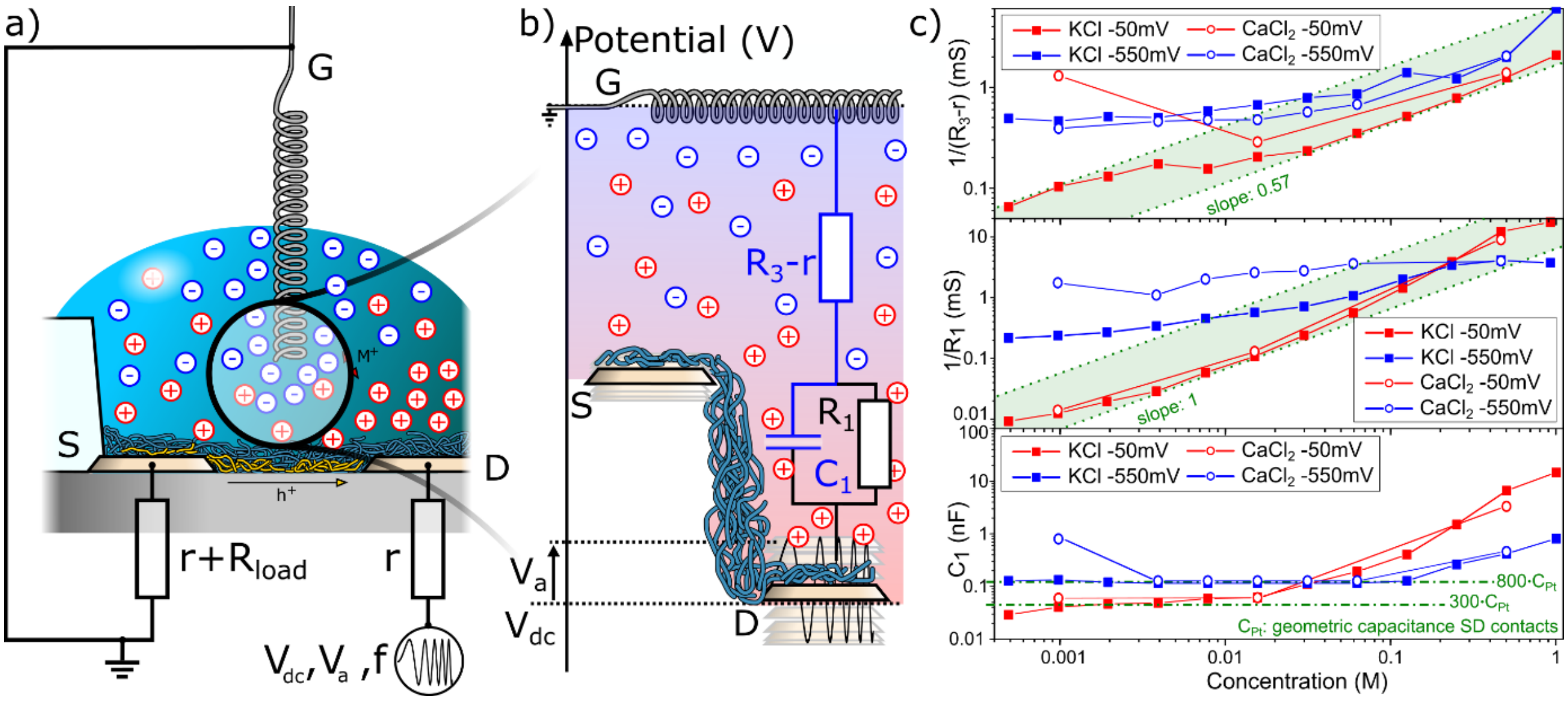

Figure 6 – (a) Experimental setup and (b) emphasis on the potential difference between the drain, the gate and the one induced on the source electrode which introduces the second RC component observed in Figure 1 by the deviation D1 to the 2R1C model (c) Dependency of the fitted 1/($R_3$-r), 1/$R_1$ and $C_1$ values with the concentration (parameters were fitted to the data with the model in Figure 3b).

The first deviation to the 2R1C model, characterized by the presence of $R_3$ and $C_1$ in the optimized equivalent circuit and its reduced form Model 1, shows the presence of a second RC transition at higher frequencies than the characteristic frequency of the drain-gate coupling in the kHz.

Depending on the resistance of the PEDOT:PSS compared to the resistance of the load, the source potential modulates the contribution of the drain electrode in the signal transmission through the electrolyte: acting either as an extended drain (in case the PEDOT is doped) or as an extended gate (in case the PEDOT is dedoped - Figure 6a). Having distinctive potentials, the three electrodes are involved in three capacitive coupling through the electrolyte from which, the vicinity of the drain with the source induces the highest capacitance compared to the coupling of the gate wire with any of the source or the drain electrodes (Figure 6b). The independent ion accumulation at both source and drain coupled electrodes ($C_1$) also induces a local ionic transport ($R_1$) in addition to the transport ($R_3$-r) from the gate wire to the micro-fabricated OECT device.

In Figure 6c, one can notice that both $R_1$ and $R_3$ decrease with the concentration for both $KCl_{(aq)}$ and $CaCl_{2(aq)}$ electrolytes at both -50 mV and -500 mV polarizations. These systematic decreases confirm the electrolytic nature of these resistors (for which the conductivity increases with the cation concentration) opposed the behavior of electronic ones (for which the conductivity decreases with the cation-supported polymer dedoping).

Also, we noticed that in all four conditions of electrolyte and voltage, the monotonicity of the concentration of these resistances seems more controlled for $R_1$ than for $R_3$ which might be explained by the lack of systematic experimental reproducibility for introducing the macro gate in the electrolyte droplet compared to the controlled distance between the lithographically patterned source and drain electrodes. For concentrations lower than 0.25 M, both resistances decrease with $V_{DC}$. Surprisingly, the resistance $R_3$ does not seem highly cation dependent, while a cation dependency is confirmed for $R_1$ upon high $V_{DC}$ polarization and for concentrations below 0.25 M (with $CaCl_{2(aq)}$ electrolytes being more conductive than $KCl_{(aq)}$ ones, confirming the ion drift mobility dependency).[9,17]

The dependency of $R_1$ with the concentration, confirming its electrolytic nature, also assesses the nature of its parallel $C_1$ to be a geometric capacitor rather than a double layer capacitor (rather associated to a charge-transfer resistance). $C_1$ does not show apparent cation dependency: At concentration below $10^{-1}$ M, $C_1$ has low capacitance at the range from 30 to 110 pF with almost no dependence on the concentration. At higher concentrations, an increase of $C_1$ appears, which might be attributed to modification of the OECT potential profile caused by dedoping. Its DC voltage dependency could also be justified by the higher dedoping for PEDOT:PSS. From the analytical expression for the geometric capacitance for concentric coplanar sensors of Cheng and coworkers,[18] we estimated the contribution "$C_{Pt}$" of the naked source and drain electrode to be only 0.15 pF: which is two orders of magnitude higher than the low concentration plateau we observed in figure 5c. This means that the polarization of PEDOT:PSS has a main contribution in this geometric capacitance $C_1$ through the electrolyte.

Overall, we can notice that although the usage of a large macro-gate wire made of a fast redox-couple material promotes the drain-gate voltage drop through the resistive electrolyte and not at the gate electrodes interface,[19] there is a competitive local gating effect originating from the source and its difference of polarity with the drain. Being inherent to the OECT operation as a transistor to extract the PEDOT:PSS conductance, the competitive self-gating of the device shall be taken into account with a potential capacitive coupling originating from the source and drain electrodes vicinity. Such "local gating" effects through the electrolyte shall also arise in the case of OECT arrays, as shown by Gkoupidenis and coworkers, through the inter OECT communication at millimeters distances.[20]

### 4.2 The Second Deviation - The Inductive Contribution

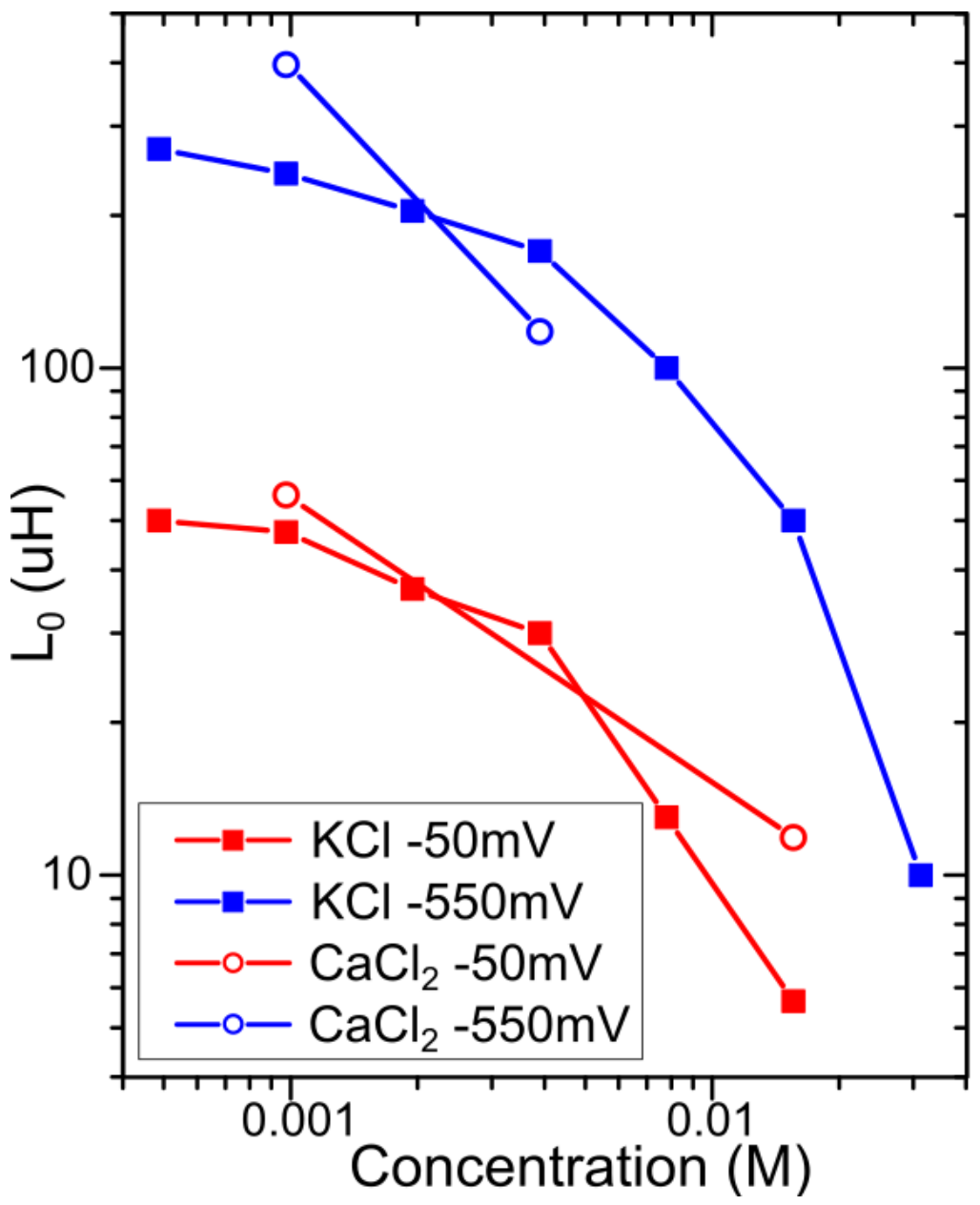

Figure 7 –Dependency of the fitted $L_0$ values with the concentration (parameters were fitted to the data with the model in Figure 3b).

To justify the positive reactance observed at high frequencies, we introduced an inductor ($L_0$) as inductive element in our model. Trends in the induction behavior are monotonic and reproducible from one ion series to the other. Fitting converged with $L_0$ values at the order of 10-100 µH for both cation but exclusively for concentrations sufficiently low (typically below $3{\cdot}10^{-2}$ M, depending on the cation and $V_{DC}$), increasing with the decrease of the ion concentration (see also Figures S1-S3 in supplementary materials). As mentioned before at high concentrations, it is not possible to fit $L_0$ due to missing positive reactance at these concentrations, and $L_0$ is set to $L_0 = 0$. $L_0$ values depend on the applied voltage $V_{DC}$ but do not seem to be significantly depending on the nature of the electrolyte. Positive reactance at high frequencies has often been observed in various electrochemical systems,[21-24] with its origin being often quite controversial.[25,26] Main hypothesis on the empirical observation of such induction at high frequencies are (i) spectrometer artefacts, (ii) cable current loops, (iii) electrodes geometry and electrochemical events such as (vi) solute adsorption-desorption dynamics.[26] From the analytical approximation of Gao and workers,[27] the inductance of the outer-ring source electrode is at maximum four orders of magnitude lower than the lowest fitted values $L_0$ displayed in figure 7: the geometry of the source electrode does not seem to be the origin of the apparent inductance. The presence of an inductor in PEDOT:PSS-based OECT channels was already proposed by Tu and coworkers to explain an empirically observed self-oscillation of the device at a resonance of 10 Hz, caused by an apparent inductor of 3 H.[28] But considering the difference in frequency range and in inductance value, with respect to the different device geometry and materials, a common origin for both inductive effects appears as unlikely (we want to mention that low frequencies, inductions have also been characterized in electrochemical systems and attributed to different mechanisms, such as corrosion).[29,30] Anyhow, the dependence of inductance on the cation concentration in our device might suggest an electronic current modulation upon device polarization due to the doping/dedoping process driven by the applied harmonic voltage as a underlying mechanism for positive reactance. As such, this positive reactance could be an inherent property of the OECT devices. The detailed analysis will be considered in another study.[31]

### 4.3 The Third Deviation – The Slow Dynamics Contribution

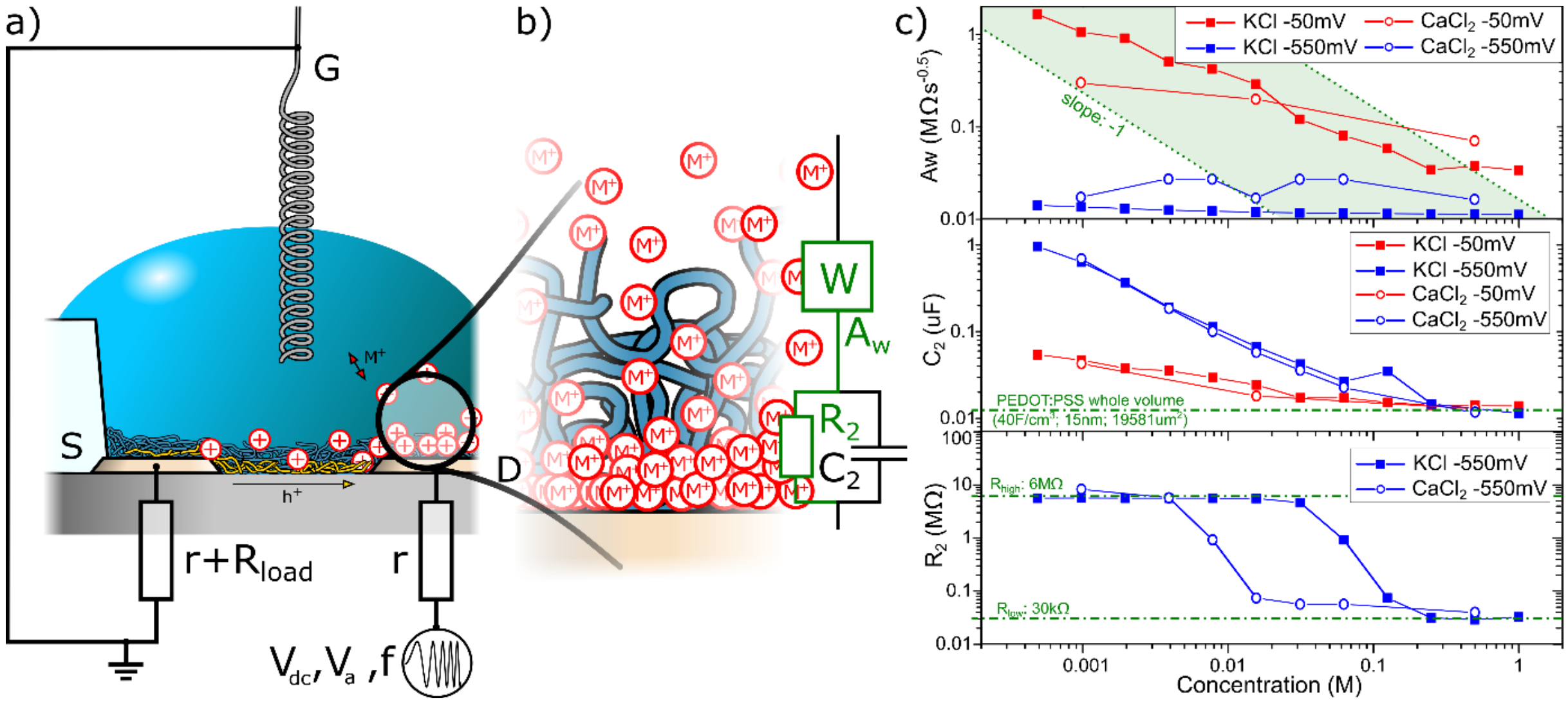

Figure 8 - (a) Experimental setup and (b) emphasis on the impedance of the drain contacts, considering both Warburg and capacitive charges. (c) Dependency of the fitted Aw, $C_2$ and $R_2$ values with the concentration (parameters were fitted to the data with the model in Figure 3b).

We observed the Warburg coefficient to be highly depending on the applied voltage (Figure 8c). The Warburg element almost does not express at -550 mV with respect to the case of applied -50 mV, where we observe a higher coefficient: up to two orders of magnitude for the lowest $KCl_{(aq)}$ concentration. At -50 mV, one can observe also a decrease of the Warburg coefficient with the bulk concentration, as predicts its analytical expression based on diffusion limitation (noticing a deviation from this law, considering slopes higher than -1 in the double logarithmic graph).[32] The Warburg coefficient does not show substantial dependency on the nature of the cation. Our motivation for choosing a Warburg element to model the non-ideal charging behavior originated from the high water permeability of this conducting polymer. It is the main feature which distinguishes OECTs from other electrolyte-gated organic field-effect transistors. For the charging properties, Rivnay and coworkers have shown by impedance spectroscopy that the total charge OECTs can store is rather function of the volume of PEDOT:PSS than its surface (with a volume capacity of 40 $F{\cdot}cm^{-3}$) implying PEDOT to be involved in the double layer formation with cations.[11] For the transport properties, Stavrinidou and coworkers experimentally characterized transient charge distribution in polarized PEDOT:PSS and modeled the time-dependent voltage-drop across the polymer by a serial distribution of two resistors.[33,34] As a transmission line of both distributed series electrolytic resistors, with local capacitances in parallel at all levels of the polymer, the Warburg element (associated to a coefficient Aw) appears to be a constant phase element suitable, for the infinite resistive/capacitive-elements discretization for the impedance of ion transported-through/charging the PEDOT:PSS.[35] In the optimized model (Figure 3b), the association of the Warburg in series with a capacitor translates two successive and distinctive charging modes rather than a single multi-processes charge (like in a Randle circuit for instance). Therefore, the accumulation at the interface of the electrode has been considered, as a parallel resistor capacitor circuit ($R_2$ and $C_2$). Considering a potential contact resistance at the metal/semiconductor interface (in practice, we observed that devices with Pt electrodes performed an order of magnitude lower than with Au), and the thickness of our PEDOT:PSS (considering ref [34], the front of the cation charges accumulation upon device polarization shall reach the drain electrode in about 0.2 µs at $V_{DC}$ = -550 mV), it is legitimate to expect, in the frequency range of our measurements, some charge accumulation also at the interface of the metal with the electrolyte channels in the polymer. In the end, the impedance of the PEDOT:PSS-coated electrode resemble to the Randles circuit, at the specificity that the double-layer forms in the bulk of the polymer in series with the diffusion element, rather than on its surface in parallel of the diffusion-limited transport (Figure 8a and 8b). Giving the swelling capability of PEDOT:PSS in water to be of 55%,[33] the model suggesting ions being able to penetrate the layer seems totally adequate.

The $C_2$ contribution is the major capacitive contribution in our system (being up to 3-fold higher than $C_1$ for the lowest concentrations). In Figure 8c, one can observe a very smooth tendency to decrease over two orders of magnitude with the concentration. Also, $C_2$ seems to be higher upon higher $V_{DC}$ polarization, but do not show any specificity on the nature of the cation. Considering our geometries, the obtained $C_2$ capacitances are much higher than we expected. First we noticed, considering the volume capacitance of PEDOT:PSS,[11] that the lowest values for $C_2$ obtained for concentrations above 0.125 M correspond to an integral charge of the whole of PEDOT:PSS volume (volume above the drain accounting for 42%, above the channel 39% and above the source 19% of this whole volume). This results are in line with other studies showing that the performances of large contacts OECTs depend rather on the whole volume of polymer than the specific volume in the channel region.[36] Within the specific frequency range for our device, this shows that the steady double layer forms on the whole PEDOT:PSS, characterized by $C_2$. On the justification for the further increase of $C_2$ by decreasing the concentration up to 1 µF, it appears not plausible to attribute it to our micro-fabricated OECT device alone, giving the fact that the lowest values at high concentration correspond to the involvement of the entire PEDOT:PSS material for the double-layer capacitance. One might attribute the higher double-layer capacitance to the participation of the double-layer at the gate electrode, involved in another 1R1C component characteristic to the gate/electrolyte interface, which has not been taken into account in our optimized model, and often neglected under specific conditions:[37] The condition for neglecting the interfacial impedance of the gate with respect to the device is to have (i) a negligible charge-transfer resistance at the gate (here using Ag/AgCl as a fast redox couple at the gate) and (ii) the highest gate/PEDOT:PSS surface ratio (here using of a macroscopic gate).[37] Since by decreasing the concentration of ion, we also decrease specifically the concentration of chloride ions participating in the charge transfer of the gate (which operates under oxidation mode upon $V_{DC}$ polarization), the decrease of concentration also affects the expression of the double-layer gate capacitance.

The resistance $R_2$ is the one responsible of the electrolyte path change under low frequencies and the associated deviation D3. Its presence has therefore an impact on the whole frequency range of the impedance spectrograms. It corresponds to the only electrolytic current transport which is not originating from a serial capacitive coupling in the equivalent circuit, therefore the Faradaic conversion of an ionic current into an electronic one: it is a charge-transfer resistance. At $V_{DC}$ = -50 mV, its value is very high (the fittings did not converge to values lower than 10 GΩ) and the element can be neglected to simplify the model without affecting the other fitted parameters. But at $V_{DC}$ = -550 mV for both $KCl_{(aq)}$ and $CaCl_{2(aq)}$, the charge-transfer resistance $R_2$ tends to switch between two states: from $R_{low}$ = 30 kΩ at high concentration to $R_{high}$ = 6 MΩ at low concentration. While the nature of the reduction(s) involved in the process of $R_2$ has (or have) not been clearly identified yet, it is worth noticing this change between two resistive states to be abrupt (two decades in resistance for one decade in concentration) and seems to be cation dependent. From figure 8c, it appears that, when polarized at -550 mV, the reduction switch occurs on the device with ten-time fewer $Ca^{2+}_{(aq)}$ than $K^{+}_{(aq)}$. A possible explanation could lie on a same Faradaic conversion on two different sites, either at the Pt surface or in the PEDOT:PSS, for which the preferential site is modulated by the dedoping level of the PEDOT:PSS. This charge-transfer resistance being cation specific and considering the associated redox-driven mechanism to be completely uncorrelated from the PEDOT:PSS conductivity nor the electrolyte one, perspectives to exploit this feature in OECTs as a higher-dimensional multi-parametric sensor are encouraging.

## Conclusion

The model of our study definitely concludes on the multi-parametric character of the OECT as an ion sensor and the complexity of its associated impedance. We pointed out from the impedance spectroscopy of our OECT three main deviations at different electrolytic conditions which corresponds to specific limitations, allowing to locally probe the OECT at different levels to assign individual impedance contributions over the whole OECT architecture. We proposed an optimized common model for any electrolytes: gathering electronic properties of the device with non-ideal electrochemical impedances of the rich PEDOT:PSS material. From this model, we demonstrated that several discrete elements were required to model the practical frequency-dependent behavior of OECTs for various electrolytes, different in concentration and nature. Specifically, we showed from the first deviation that the gating of the polarized source induces a local ionic flow through $R_1$, competitive with the gate responsible for the global ion flow through $R_3$-r. Also from this local polarization, a minority geometric capacitance $C_1$ arises, but remains two-fold lower than the PEDOT:PSS capacitance $C_2$. We also evidenced from the last deviation a diffusion-limitation modelled with a constant phase Warburg element as well as a charge-transfer resistance $R_3$ characterizing a non-negligible Faradaic conversion under large biases. We also observed that the positive reactance of the device can be modelled by an inductor $L_0$, from which its dependence with the concentration testifies its physical origin from our electrolyte-gated device. Having pointed out further deviations to theory, increasing further the complexity of the model offers the perspective to unravel more information. Some comparisons of the fitted element values normalized with geometric specifications and literature values suggest that new elements might embed many more contributions (as for instance $C_2$ which might include gate capacitance contribution at low electrolyte concentrations). Our approach has been based on introducing the least flexibility to our model, but satisfying the data the most. And of course, increasing the model complexity might give a better picture of this many-interface device. In that scope, further work rather based on geometry variations can complement these, which are based on electrolyte and voltage variations. Pointing out also the necessity to extend our abilities to make analog microelectronic components to mimic the non-ideal impedance of neurons, as highlighted by Gutmann,[38] our study demonstrates that our micro-fabricated OECTs gather a broad range electronic properties, namely (i) resistive, (ii) capacitive, but also (iii) pseudo-capacitive (related to constant-phase elements) and (iv) inductive, for which the last two are not conventional in microelectronics but necessary to build biomimetic synaptic devices for neuromorphic sensing and neuromorphic computing.[38]

## Acknowledgement

We acknowledge financial supports from the EU: H2020 FET-OPEN project RECORD-IT (# GA 664786) and in part (RBI) by the H2020 CSA Twinning project No. 692194, RBI-T-WINNING and the European Union through the European Regional Development Fund – the Competitiveness and Cohesion

Operational Programme (KK.01.1.1.06). We thank the French National Nanofabrication Network RENATECH for financial support of the IEMN clean-room.

## References

1. Jonathan Rivnay, Sahika Inal, Alberto Salleo, Róisín M. Owens, Magnus Berggren & George G. Malliaras, Organic Electrochemical Transistors*, Nat. Rev. Mater.* **3**, 170886 (2018).

2. Sang Min Won, Enming Song, Jianing Zhao, Jinghua Li, Jonathan Rivnay & John A. Rogers, Recent Advances in Materials, Devices, and Systems for Neural Interfaces, *Adv. Mater.* **30**, 1800534 (2018).

3. Yoeri van de Burgt, Armantas Melianas, Scott Tom Keene, George Malliaras & Alberto Salleo, Organic Electronics for Neuromorphic Computing, *Nat. Rev.* **1**, 386-397 (2018).

4. Sébastien Pecqueur, Dominique Vuillaume & Fabien Alibart, Perspective: Organic electronic materials and devices for neuromorphic engineering*, J. Appl. Phys.* **124**, 151902 (2018).

5. Jia Sun, Ying Fu & Qing Wan, Organic Synaptic Devices for Neuromorphic Systems, *J. Phys. D: Appl. Phys.* **51**, 314004 (2018).

6. Dion Khodagholy, Thomas Doublet, Pascale Quilichini, Moshe Gurfinkel, Pierre Leleux, Antoine Ghestem, Esma Ismailova, Thierry Hervé, Sébastien Sanaur, Christophe Bernard & George G. Malliaras, In vivo recordings of brain activity using organic transistors, *Nat. Commun.* **4**, 1575 (2013).

7. Paschalis Gkoupidenis, Nathan Schaefer, Benjamin Garlan & George G. Malliaras, Neuromorphic Functions in PEDOT:PSS Organic Electrochemical Transistors, *Adv. Mater.* **27**, 7176-7180 (2015).

8. Sébastien Pecqueur, Maurizio Mastropasqua Talamo, David Guerin, Philippe Blanchard, Jean Roncali, Dominique Vuillaume, Fabien Alibart, Neuromorphic Time-Dependent Pattern Classification with Organic Electrochemical Transistor Arrays, *Adv. Electron. Mater.* **4**(9), 1800166 (2018).

9. Sébastien Pecqueur, David Guérin, Dominique Vuillaume & Fabien Alibart, Cation Discrimination in Organic Electrochemical Transistors by Dual Frequency Sensing, *Org. Electron.* **57**, 232-238 (2018).

10. Bernards, D. A.; Malliaras, G. G. *Adv. Func. Mater.* **17**, 3538-3544, (2007).

11. Rivnay, J.; Leleux, P.; Ferro, M.; Sessolo, M.; Williamson, A.; Koutsouras, D. A.; Khodagholy, D.; Ramuz, M.; Strakosas, X.; Owens, R. M.; Benar, C.; Badier, J.-M.; Bernard, C.; Malliaras, G. G. *Science Advances* **1**, e1400251 (2015).

12. Worfolk, Brian J.; Andrews, Sean C.; Park, Steve; Reinspach, Julia; Liu, Nan; Toney, Michael F.; Mannsfeld, Stefan C. B.; Bao, Zhenan. *"Ultrahigh electrical conductivity in solution-sheared polymeric transparent films"*. Proceedings of the National Academy of Sciences. **112** 46), 14138–14143 (2015).

13. M. A. Branch, T. F. Coleman, and Y. Li, "A Subspace, Interior, and Conjugate Gradient Method for Large-Scale Bound-Constrained Minimization Problems," SIAM Journal on Scientific Computing **21**(1), 1-23 (1999).

14. Jones E, Oliphant E, Peterson P, et al. SciPy: Open Source Scientific Tools for Python, 2001-, http://www.scipy.org

15. Vikash Kaphle, Shiyi Liu, Akram Al-Shadeedi, Chang-Min Keum, Björn Lüssem, Contact Resistance Effects in Highly Doped Organic Electrochemical Transistors, *Adv. Mater.*, **28**(39), 8766-8770 (2016).

16. Jonathan Rivnay, Sahika Inal, Brian A. Collins, Michele Sessolo, Eleni Stavrinidou, Xenofon Strakosas, Christopher Tassone, Dean M. Delongchamp, and George G. Malliaras, Structural control of mixed ionic and electronic transport in conducting polymers, *Nat. Commun.* **7**, 11287 (2016)

17. Nicolas Coppedè, Marco Villani Francesco Gentille, Diffusion Driven Selectivity in Organic Electrochemical Transistors, *Sci. Rep.* **4**, 4297 (2014)

18. Huanyu Cheng, Yihui Zhang, Xian Huang, John A. Rogers, Yonggang Huang, Analysis of a concentric coplanar capacitor for epidermal hydration sensing, *Sens. Actuators, A* **203**, 149-153 (2013)

19. Tarabella, G.; Santato, C.; Yang, S.Y.; Iannotta, S.; Malliaras, G.G.; Cicoira, F. Effect of the gate electrode on the response of organic electrochemical transistors. *Appl. Phys. Lett.* **97**, 123304 (2010).

20. Paschalis Gkoupidenis, Dimitrios A Koutsouras & Georges G Malliaras, Neuromorphic device architectures with global connectivity through electrolyte gating, *Nat. Commun.* **8**, 15448 (2017).

21. Shalini Rodrigues N. Munichandraiah A. K. Shukla, AC impedance and state-of-charge analysis of a sealed lithium-ion rechargeable battery, *J. Solid State Electrochem.*, **3**(7-8), 397-405 (1999).

22. N.Wagner, W.Schnurnberger, B.Müller & M.Lang, Electrochemical impedance spectra of solid-oxide fuel cells and polymer membrane fuel cells, *Electrochim. Acta* **43**(24), 3785-3793 (1998).

23. Haifeng Dai, Bo Jiang & Xuezhe Wei, Impedance Characterization and Modeling of Lithium-Ion Batteries Considering the Internal Temperature Gradient, *Energies* **11**, 220 (2018).

24. Tetsuya Osaka, Daikichi Mukoyama and Hiroki Nara, Review—Development of Diagnostic Process for Commercially Available Batteries, Especially Lithium Ion Battery, by Electrochemical Impedance Spectroscopy, *J. Electrochem. Soc.* **162**(14), A2529-A2537 (2015).

25. Michel Keddam, Christiane Rakotomavo, Hisasi Takenouti, Impedance of a porous electrode with an axial gradient of concentration, *J. Appl. Electrochem.* **14**(4), 437-448 (1984).

26. P. Córdoba-Torres, M. Keddam, R. P. Nogueira, On the intrinsic electrochemical nature of the inductance in EIS: A Monte Carlo simulation of the two-consecutive-step mechanism: The flat surface 2 D case, *Electrochim. Acta* **54**(2), 518-523 (2008).

27. Wei Gao, Chao Jiao, Zhiping Yu, “Efficient inductance calculation for planar spiral inductors and transformers based on analytical concentric half-turn formulas”, *Int. J. RF Microw. Comput.-aided Eng.* **16**(6), 565-572 (2006).

28. Deyu Tu, Robert Forchheimer, Self-oscillation in electrochemical transistors: An RLC modeling approach, *Solid-State Electron.* **69**, 7-10 (2012).

29. Nandini Dinodi, A. Nityananda Shetty, Electrochemical investigations on the corrosion behaviour of magnesium alloy ZE41 in a combined medium of chloride and sulphate, *J. Magnesium Alloys* **1**(3), 201-209 (2013).

30. A. D. King, N.Birbilis, J. R. Scully, Accurate Electrochemical Measurement of Magnesium Corrosion Rates; a Combined Impedance, Mass-Loss and Hydrogen Collection Study, *Electrochim. Acta* **121**, 394-406 (2014).

31. Crljen et al., *to be published*

32. Cynthia G. Zoski, Handbook of Electrochemistry (1$^{st}$ ed) Elsevier 2007 pp.456

33. Eleni Stavrinidou, Pierre Leleux, Harizo Rajaona, Dion Khodagholy, Jonathan Rivnay, Manfred Lindau, Sébastien Sanaur, George G. Malliaras, Direct Measurement of Ion Mobility in a Conducting Polymer, *Adv. Mater.* **25**(32), 4488-4493 (2013).

34. Eleni Stavrinidou, Pierre Leleux, Harizo Rajaona, Michel Fiocchi, Sébastien Sanaur, and George G. Malliaras, A simple model for ion injection and transport in conducting polymers, *J. Appl. Phys.* **113**, 244501 (2013).

35. J C Wang, Realizations of Generalized Warburg Impedance with RC Ladder Networks and Transmission Lines, *J. Electrochem. Soc.* **134**(8), 1915—1920 (1987).

36. Pecqueur S, Lenfant S, Guérin D, Alibart F, Vuillaume D, “Concentric-electrode organic electrochemical transistors: case study for selective hydrazine sensing”, *Sensors* **17**(3), 570 (2017).

37. Fabio Cicoira, Michele Sessolo, Omid Yaghmazadeh, John A. DeFranco, Sang Yoon Yang, and George G. Malliaras, Influence of Device Geometry on Sensor Characteristics of Planar Organic Electrochemical Transistors, *Adv. Mater.* **22**, 1012-1016 (2010).

38. F. Gutmann, Electrochemical Pseudo-Inductances, *J. Electrochem. Soc.* **112**(1), 94-98 (1965).

# Supporting Information for:

# The Non-Ideal Organic Electrochemical Transistors Impedance

**Sébastien Pecqueur[1,§,*], Ivor Lončarić[2,§], Vinko Zlatić[2], Dominique Vuillaume[1], Željko Crljen[2,*]**

[1] Institute of Electronics, Microelectronics and Nanotechnologies (IEMN), CNRS, Univ. Lille, Villeneuve d'Ascq, France
[2] Institut Ruđer Bošković, P.O. Box 180, 10002 Zagreb, Croatia
[§] these authors have contributed equally
* E-mail: sebastien.pecqueur@*iemn.univ-lille1*.fr; zeljko.crljen@*irb*.hr

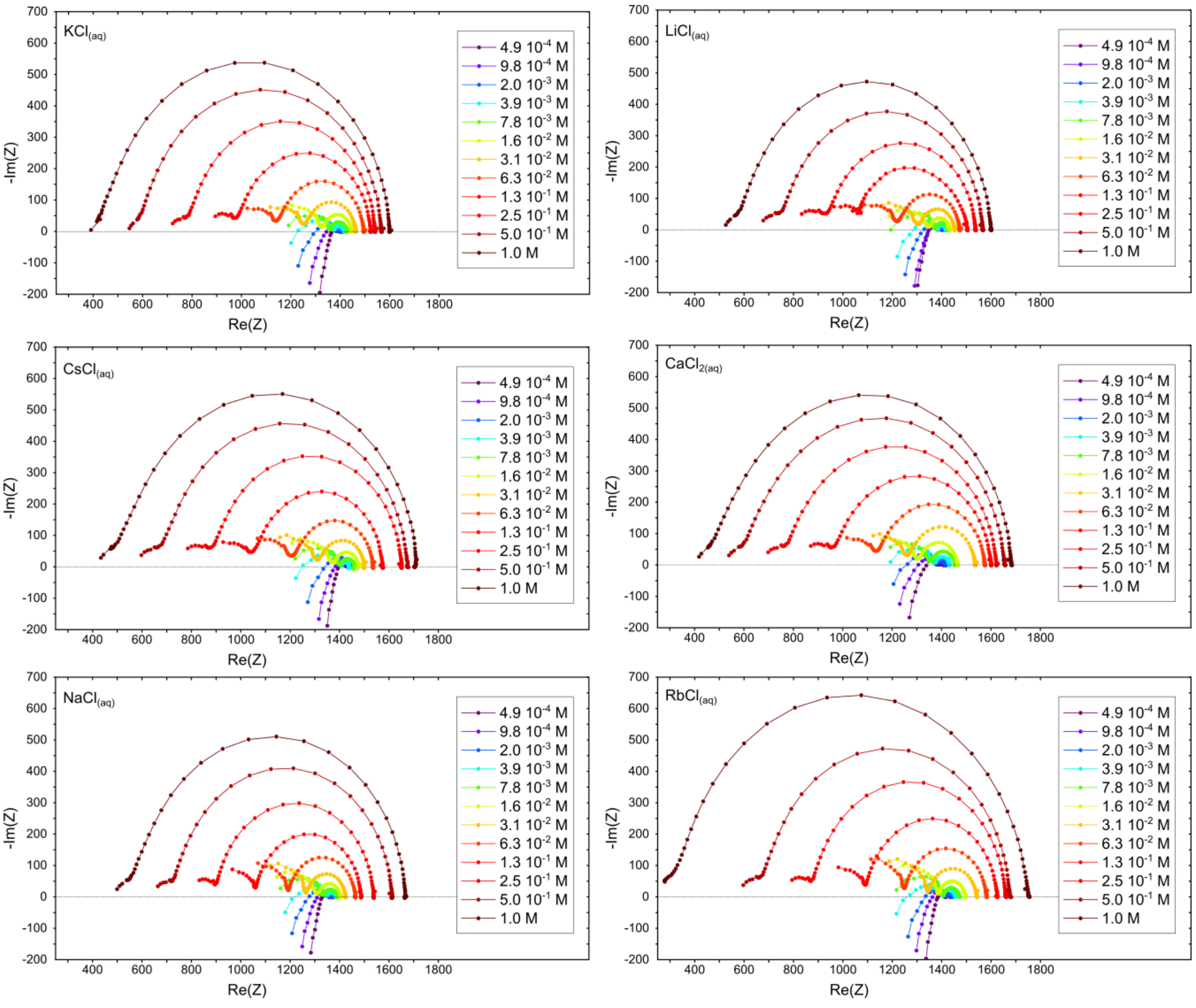

Figure S1 – Nyquist plots for the OECT device, measured in six different cation aqueous electrolytes at the concentration $1/2^n$ M (n from 0 to 11) at a stress of DC level = -50 mV and amplitude 50 mV at the drain, 1 kΩ-loaded source bridged to grounded gate.

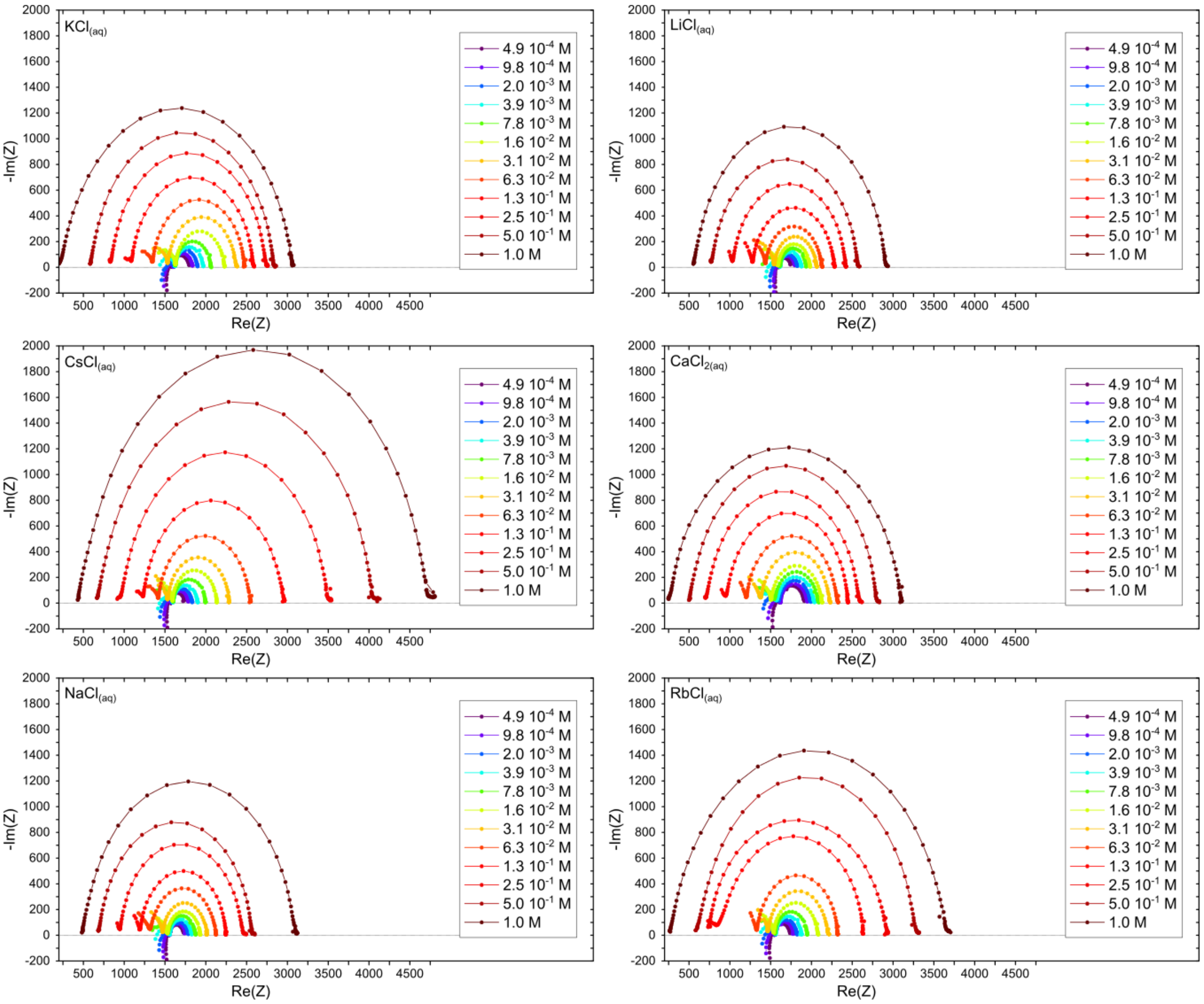

Figure S2 – Nyquist plots for the OECT device, measured in six different cation aqueous electrolytes at the concentration $1/2^n$ M (n from 0 to 11) at a stress of DC level = -350 mV and amplitude 50 mV at the drain, 1 kΩ-loaded source bridged to grounded gate.

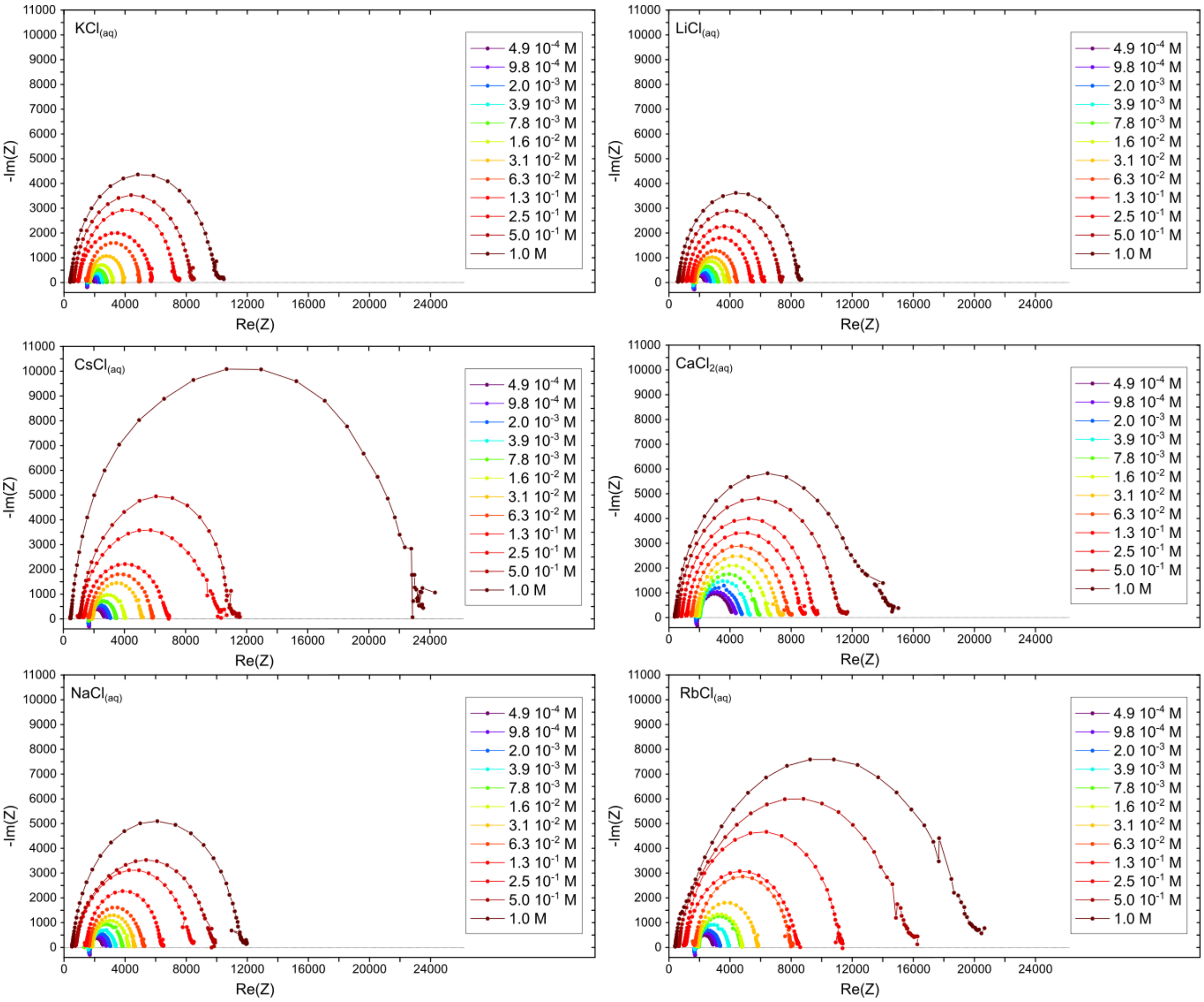

Figure S3 – Nyquist plots for the OECT device, measured in six different cation aqueous electrolytes at the concentration $1/2^n$ M (n from 0 to 11) at a stress of DC level = -550 mV and amplitude 50 mV at the drain, 1 kΩ-loaded source bridged to grounded gate.

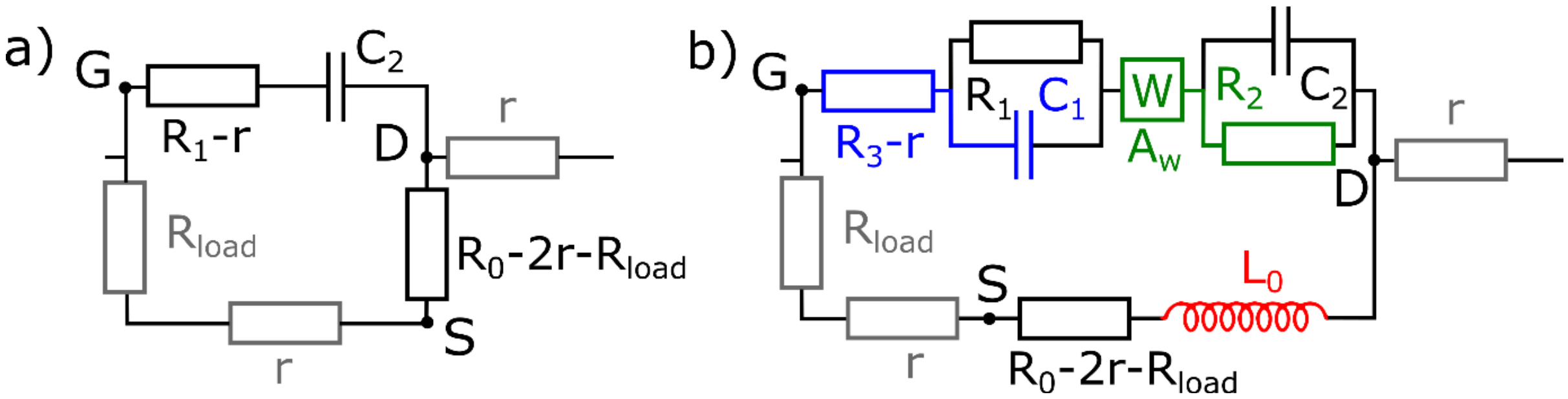

Figure S4 – Transformation of the original "2R1C" model (Figure 3a) and the optimized model (Figure 3b), by integrating the resistive elements (r and $R_{load}$) of the setup environment (these elements are no variable and are not increasing the flexibility of the models).

*We want to highlight that the equivalence between models displayed in Figures 3a and S4a is acceptable if $r^2 << R_0R_1$ and $2r << R_0+R_1$. Considering the low resistivity of the contacts compared to the resistances of OECT/electrolyte system, the transformation is legitimate within its usage for the data which were obtained in this study (applies also to Figure 3b and S4b).*